\documentclass[%
 reprint,
superscriptaddress,
 amsmath,amssymb,
aps,
floatfix
]{revtex4-2}
\usepackage{amsmath, amsthm, amssymb}
\usepackage[pdftex]{graphicx}
\usepackage{dcolumn} % Align table columns on decimal point
\usepackage{xcolor}
\usepackage{MnSymbol,wasysym}

\usepackage{bm} % bold math
\usepackage{hyperref}
\usepackage[normalem]{ulem}

\newcommand{\HH}{\mathcal{H}}
\newcommand{\vk}{{\bf k}}
\newcommand{\vp}{{\bf p}}

\newcommand{\vrr}{{\bf r}}

\newcommand{\pdag}{\phantom{\dag}}

\begin{document}

\title{A mechanism for  $\pi$ phase shifts in Little-Parks experiments: application to 4Hb-TaS$_2$ and to 2H-TaS$_2$ intercalated with chiral molecules}
\author{Mark H. Fischer}
 \affiliation{%
 Department of Physics, University of Zurich, Winterthurerstrasse 190, 8057 Zurich, Switzerland
 }%
 \author{Patrick A. Lee}
\affiliation{%
Department of Physics, Massachusetts Institute of Technology, Cambridge, MA 02139 USA
}%
 \author{Jonathan Ruhman}
 \affiliation{%
 Department of Physics, Bar Ilan University, Ramat Gan 5290002, Israel
 }%
 
%\date{2023}

\begin{abstract}
    Recently, unusual $\pi$ phase shifts in Little-Parks experiments performed on  two systems derived from the layered superconductor 2H-TaS$_2$ were reported. These systems share the common feature that additional layers have been inserted between the 1H-TaS$_2$ layers. 
    In both cases, the $\pi$ phase shift has been interpreted as evidence for the emergence of exotic superconductivity in the 1H layers.
    Here, we propose an alternative explanation assuming that superconductivity in the individual 1H layers is of conventional $s$-wave nature derived from the parent 2H-TaS$_2$. We show that a negative Josephson coupling between otherwise decoupled neighboring 1H layers can explain the observations. Furthermore, we find that the negative coupling can arise naturally assuming a tunneling barrier containing paramagnetic impurities. An important ingredient is the suppression of non-spin-flip tunneling due to  spin-momentum locking of Ising type in a single 1H layer together with the inversion symmetry of the double layer. 
    In the exotic superconductivity scenario, it is challenging to explain why the critical temperature is almost the same as in the parent material and, in the 4Hb case, the superconductivity's  robustness  to disorder. Both are non-issues in our picture, which also exposes the common features that are  special in these two systems.
\end{abstract}

\maketitle

\begin{figure}[tt]
\centering
\includegraphics[width=3.25in]{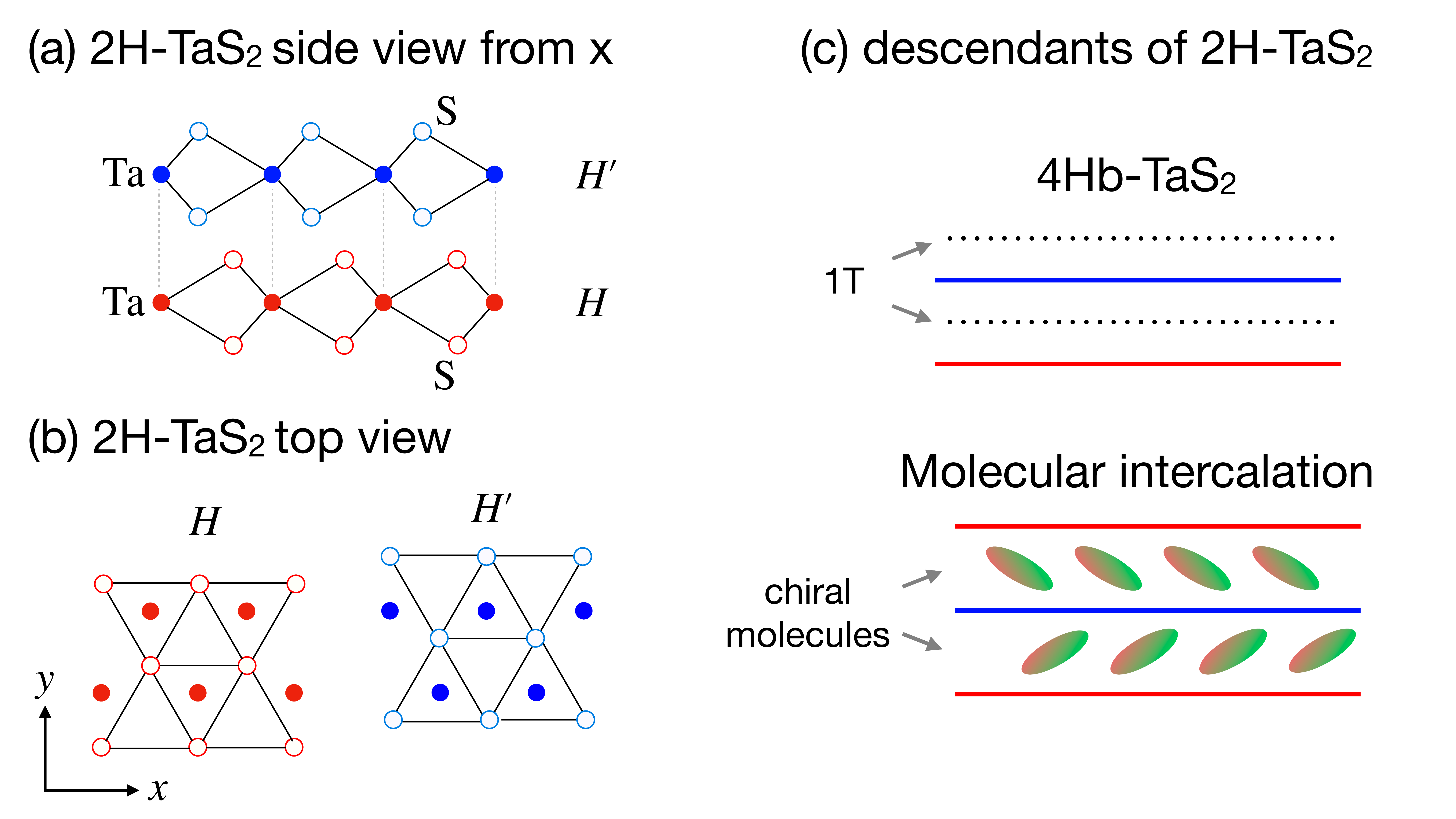}
\caption{(a) Side view of the crystal structure of 2H-TaS$_2$. The full circles are Ta atoms while empty circles are S. The unit cell consists of two layers that are inversion partners, denoted by $H$ (red) and $H'$ (blue). (b) Top view of the two types of H-layers. 
(c) Types of descendants of 2H-TaS$_2$ considered here. Top: 4Hb-TaS$_2$, where monolayers of 1T-TaS$_2$ separate the $H$ and $H'$ layers. In 1T, the S layer above Ta is rotated by  $60^{\circ}$ relative to the layer below. 
Bottom: chiral molecules intercalated between the 1H layers. 
}
\label{fig:crystal structure}
\end{figure}

\emph{Introduction---}A $\pi$ phase shift in Little-Parks experiments 
is usually taken as a strong indication for unconventional superconductivity~\cite{geshkenbein1987vortices, li:2019a, xu:2020}.
A recent experiment observing such a shift involves the intercalation of a chiral molecule between  1H-TaS$_2$ layers~\cite{wan:2023tmp}, while an earlier experiment involves the compound 4Hb-TaS$_2$ which consists of alternating layers of 1T and 1H forms of TaS$_2$~\cite{almoalem:2022tmp} (see Fig. \ref{fig:crystal structure}c). In both cases, the samples were cut out of single crystals and it was proposed that the $\pi$ phase shifts can be explained by the emergence of a superconducting state with multiple degenerate order parameters, resulting in a chiral superconducting state at low temperature, instead of the conventional $s$-wave pairing that is believed to describe the pristine 2H-TaS$_2$~\cite{nagata:1992, navarro-moratalla:2016, bekaert:2020}. 
In this paper, we propose an alternative explanation of the $\pi$ phase shift that does not require the postulation of a different pairing mechanism.

Our motivation to seek a more conventional mechanism is based on the following observations:  First, in both experiments the superconducting transition temperature, $T_{\rm c}$, is hardly changed compared to other experiments in 2H-TaS$_2$~\cite{navarro-moratalla:2016, yang2018enhanced, bekaert:2020}. Second, the robustness of the superconducting state against disorder speaks  against an unconventional pairing mechanism in the 4Hb case, where data is available~\footnote{The residual resistivity of 4Hb-TaS$_2$ is of the order of $\rho\approx 65 \mu \rm{\Omega}\,\rm {cm}$~\cite{AmitPrivate}, which is 60 times greater than Sr$_2$RuO$_4$\cite{MackenzieExtremely}. Using Drude formula with a density $n=10^{19}$ cm$^{-2}$ per layer and a rough estimate for the Fermi velocity $v_F = 3\times 10^5$ m/s one reaches the conclusion that the mean-free path is of the order $\ell\approx 2$ nm. This is smaller than the coherence length which is of order $\xi\approx 20$ nm~\cite{ribak:2020}}.
The issue is even more severe in the molecular-intercalation case: we do not expect any significant charge transfer from the molecules to the TaS$_2$ layers and the effect of the molecules on the electronic structures of the metallic layers should be minimal. 
Thus, the emergence of an entirely different pairing state should be considered a great surprise. 
Finally, it is aesthetically more pleasing to find a common mechanism for both systems sharing a common parent superconductor.

In the Little-Parks experiments, the superconducting material is fabricated into a ring and the resistivity is measured as a function of the magnetic flux piercing the ring very close to the transition temperature. While the basic idea is that $T_{\rm c}$ is modulated periodically as a function of the magnetic flux, in the actual experiment the dependence of the resistivity is measured in the transition region. Importantly, for conventional superconductors, the resistivity is minimal at zero field. The $\pi$ phase-shift effect refers to cases, where the resistivity is maximal at zero field, indicating that $T_{\rm c}$ is increased and the free energy lowered by the magnetic flux~\footnote{Note that 0 and $\pi$ phase shifts are both allowed when time-reversal symmetry (TRS) is preserved, while broken TRS in principle allows for any phase.}. Such a behavior is usually taken as evidence for unconventional superconductivity and is only expected to happen in polycrystalline samples~\cite{geshkenbein1987vortices, li:2019a} or rings with weak links~\cite{shiba:1969, bulaevskii:1977}.

\emph{The $J<0$ scenario---}
We begin by stating our basic hypothesis: As seen in Fig.~\ref{fig:crystal structure}(c), in both examples, there are two 1H layers per unit cell stacked in an AB pattern. We assume that these layers are largely decoupled and inherit the conventional (intra-layer) superconductivity of the pristine TaS$_2$~\cite{yang2018enhanced}. Importantly, we argue that in the two systems, where the $\pi$ shift was observed, the Josephson coupling $J$ between neighboring layers has a negative sign. Namely, $E_{\rm J} = -J \cos(\phi_l - \phi_{l+1})$ with $\phi_l$ the phase of the order parameter in layer $l$ and thus, $J<0$ favors a $\pi$ phase difference between the layers.
In what follows, we show that the combination of  negative Josephson coupling between individual 1H layers and lattice defects offers an explanation of the Little-Parks experiment without requiring a novel superconducting order parameter.

\begin{figure}[tt]
\centering
\includegraphics{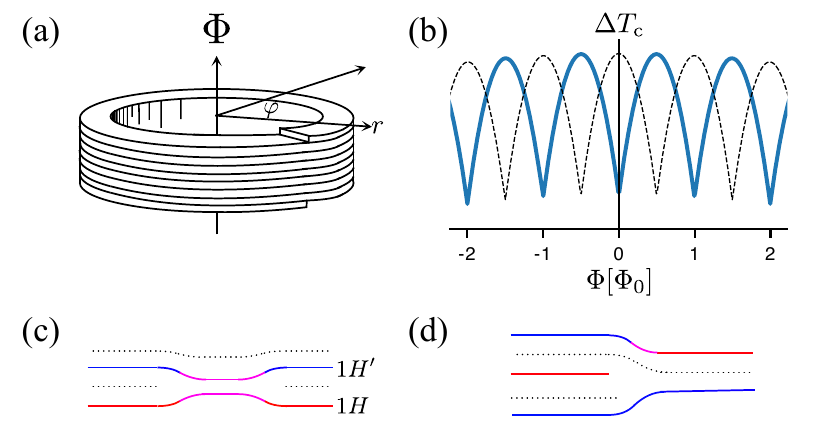}
\caption{(a) A superconducting ring made of a layered material, which encloses a single screw dislocation ($N=1$).  Due to the screw defect, a circuit around the ring is equivalent to a single-layer translation. To accommodate with the screw, the superconducting order parameter develops a $\pi$ phase shift when the layers are coupled with a negative Josephson coupling. (b) The Little-Parks oscillations of a screw-dislocation sample showing a $\pi$ phase shift (solid line) compared to the standard case (dashed line). (c) A stacking fault with a dislocation of a single layer (a missing region of 1T layer). This locally reduces the negative Josephson coupling effect. (d) A missing double layer, which necessarily incurs a $\pi$ phase shift on the 1H layers either above or below the dislocation line.}
\label{fig:dislocations}
\end{figure}

For illustration, let us first consider a ring made of a single crystal, except for a single lattice defect, a screw dislocation, that pierces the center of the ring as shown in Fig.~\ref{fig:dislocations}(a). As one completes a circuit around the ring, one ends up on another layer. The (global) pair phase is given by a slowly varying function multiplied by a sign,  $\phi(\vrr) = (-1)^l \phi(r, \varphi)$, which switches from one layer to the next due to the negative Josephson coupling. Here, $r$ is the radial coordinate and $\varphi$ the in-plane angle. The screw dislocation therefore creates a phase mismatch of $\pi$. In order to smoothly connect the layers after going around once, the phase $\phi(r, \varphi)$ must wind by $\pi$ costing kinetic energy. In contrast, applying a magnetic field yielding a half flux quantum through the ring requires no such phase winding and the associated energy cost, see App.~\ref{app:GL}. 
This produces the $\pi$ phase shifts in the Little-Parks experiment as shown in Fig.~\ref{fig:dislocations} (b).

In general, we can expect that the screw dislocation will be frozen in during the deposition of the thin film. Furthermore, it is likely that the winding number N is not unity, but can be even or odd. 
Odd N favors a $\pi$ phase shift while even N favors zero phase shift. 
There is a small complication in the systems under consideration because each unit cell contains two 1H layers, labelled $H$ and $H'$ in Fig.~\ref{fig:crystal structure}. Therefore, N odd technically corresponds to a half-integer screw dislocation,
which will necessarily induce a domain-wall boundary between the $H$ and $H'$ layer.
%{\color{red} \sout{ when viewed from the top, one type has sulphur atoms forming up triangles, while the other has down triangles (see Fig.~\ref{fig:crystal structure}b).}}
We will assume that the domain wall energy is small and not sufficient to form a bias between the integer and half-integer screw dislocations.

The above scenario of a nearly perfect single crystal is of course unrealistic. Indeed, it is known that transition metal dichalcogenides (TMDs) are prone to developing stacking faults. We can include these faults by decorating the screw dislocation with additional edge dislocation lines that lie in the plane. We show some examples in Fig.~\ref{fig:dislocations}. However, the main effect of these additional dislocations will be a reduction of the energy splitting between zero and $\pi$ phase winding.
For a ring with a general structural defect, we expect the phase $\phi(\vrr)$ will adjust in a slowly varying way to minimize the kinetic energy. Inserting a half flux quantum through the ring again leads to a re-adjustment of the overall phase. However, due to the built in frustration arising from the sign change between layers, we can expect that in general the state with zero flux may have a free energy larger or smaller than the state with half flux quantum, depending on the detailed defect structure.
Note that a key prediction of this picture is that on average about half the samples will show a $\pi$ phase shift, while the other half show no phase shift. This is indeed consistent with the reports on 4Hb-TaS$_2$~\cite{almoalem:2022tmp}.

It should be pointed out that our picture is quite general and does not rely on the fact that the unit cell of 2H-TaS$_2$ contains two superconducting layers. However, as we will see in the following, the specific symmetry of the 2H system, with its two inversion-broken layers connected by inversion symmetry~\cite{fischer:2023}, greatly enhances the chances of finding this effect.

For the remainder of this work, we provide explanations for the origin of the postulated negative Josephson coupling between neighboring 1H layers.
We base our discussion on an early paper by Kulik~\cite{kulik:1966}, who considered the presence of a  spin-flip tunnelling amplitude $t_{\rm sf}$ due to paramagnetic impurities in the tunneling barrier in addition to the standard spin-independent amplitude $t_{\rm n}$. He found that the Josephson coupling has the form
\begin{equation} \label{eq:Kulik}
    J \propto (|t_{\rm n}|^2 - |t_{\rm sf}|^2).
\end{equation} 
While Kulik was interested mostly in the reduction of the Josephson coupling in a tunneling junction, this coupling can in principle change sign, if the spin-flip term dominates~\cite{bulaevskii:1977}.
In this paper, we will extend Kulik's theory to the case, where the superconductors have strong spin-orbit coupling (SOC), and apply it to the special case of 1H layers. 
We emphasize that in this mechanism, spin flip has to be due to scattering from magnetic impurities in the junction area, while spin flip from SOC preserves TRS and is insufficient, contrary to what was stated in Kulik's paper (see App.~\ref{app:TRS}). This is consistent with more recent papers that focus on the magnetic-impurity case~\cite{shiba:1969, bulaevskii:1977}. Even more recently, Spivak and Kivelson have emphasized the role of strong correlations, which goes beyond the effective tunneling Hamiltonian approach.~\cite{spivak:1991}.

\emph{General Formalism---}
With the individual 1H layers largely decoupled, we consider the coupling of the layers within a tunneling approach. For this purpose, we consider in the following only two 1H layers, which we denote T (top) and B (bottom). In order to discuss Kulik's mechanism in more detail and extend it to the specific case of 1H layers, we need to examine the effect of symmetries on the tunneling process as well as present some details of the band structure. We begin by reviewing the electronic structure of the 2H layer, which is the same in both cases.

\begin{figure}[tt]
\centering
\includegraphics{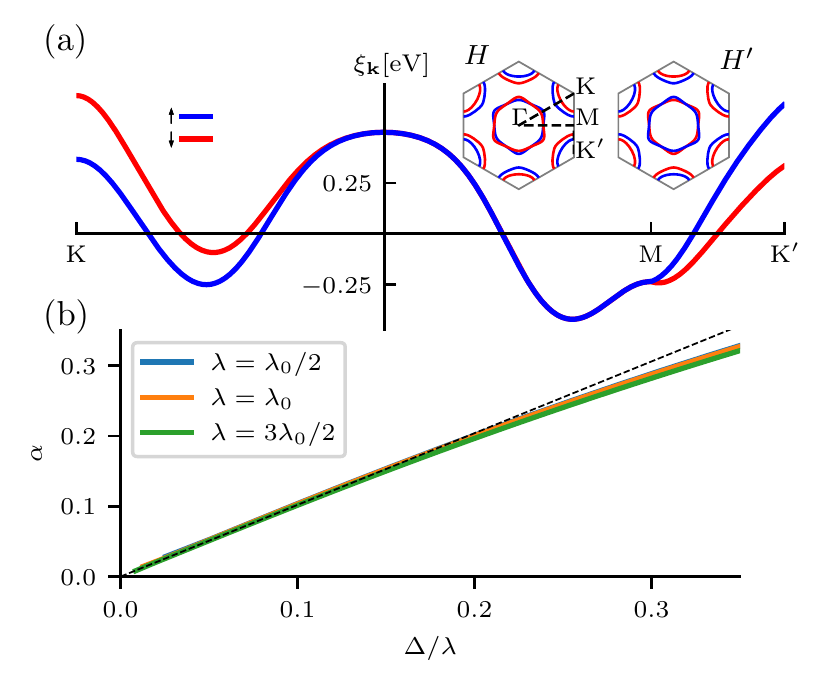}
\caption{(a) Typical 1H-TMD dispersion with the expected spin-splitting due to the Ising SOC (here $\lambda_0 \approx 0.07$ eV)~\cite{moeckli:2018, liu:2013}. The dispersion for the $H'$ layer is identical except that the spins are interchanged, as illustrated by the Fermi surfaces of the two 1H layers ($H$ and $H'$) in 2H-TaS$_2$ and its descendants such as 4Hb.
As a result, momentum conserved inter-layer tunneling for states near the Fermi surface is allowed only with spin-flip (except along $\Gamma$ to $M$). The resulting suppression of spin-conserving tunneling, denoted by $\alpha$ in Eq.~\eqref{eq:Kulik2}, is shown in panel (b) and is $\propto \Delta / \lambda$ (dashed line).}
\label{fig:bands}
\end{figure}

The individual 1H layer lacks inversion as well as $C_2$ symmetry around the $z$ axis and, as a result, has a strong (Ising) SOC. The energy band can be labeled by momentum $\vk$ and spin $s$ quantized in the $z$ direction. 
The  dispersion for a given band (e.g. in the top layer) then reads $\xi^{\rm T}_{\vk s} = \varepsilon_{\vk} + s \lambda f_\vk$ with $s=\pm$ for up and down spin, $\varepsilon_{\vk} = \varepsilon_{-\vk}$ and $f_\vk = - f_{-\vk}$. The SOC in these materials is extremely large $\lambda\sim 100$ meV~\cite{barrera:2018}. 
We see in Fig~\ref{fig:bands}(a) that this leads to a large splitting between the spin up and down bands for generic $\vk$. Importantly, note that $\lambda \mapsto -\lambda$ when going from a $H$ layer to the $H'$ layer, such that $\xi^{\rm T}_{\vk s} = \xi^{\rm B}_{\vk \bar{s}}$ with $\bar{s}$ the opposite spin. To discuss Josephson coupling, we introduce two (complex) spin-singlet $s$-wave superconducting order parameters in each layer, $\Delta^{l}_s = s|\Delta^l|e^{i\phi_l}$ with $l={\rm T}, {\rm B}$ the layer index.

For the Josephson coupling, we follow
Kulik~\cite{kulik:1966} who computed the Josephson coupling between two superconductors as the change in energy to second order in perturbation theory for a tunneling Hamiltonian
\begin{equation} \label{eq:tunnel_hopping}
    \HH_{\rm tun} = \sum_{\vk,\vp}\sum_{ss'} [T_{\vk\vp}^{ss'} \; c^\dag_{\vk,{\rm T},s} c^{\pdag}_{\vp,{\rm B}, s'} + {\rm h.c.}],
\end{equation}
where $T_{\vk\vp}^{ss'} = t_{\vk \vp}^{\rm n} \sigma^0 + t_{\vk\vp}^{\rm sf} \sigma^x$ describes both spin-independent and spin-dependent tunneling, $\sigma^0$ is the identity matrix and $\sigma^x$ is a Pauli matrix. Note that due to the origin of the spin-dependent Hamiltonian, we find $(t_{\vk\vp}^{\rm sf})^* = t_{-\vk-\vp}^{\rm sf}$. This form breaks time-reversal symmetry, which requires that $(t_{\vk\vp}^{\rm sf})^* = -t_{-\vk-\vp}^{\rm sf}$.
In contrast to ~\cite{kulik:1966}, we include the spin index to label the states, as this is crucial for our discussion. The spin-independent and spin-dependent corrections read
\begin{align}
    \Delta E_{\rm n} &= - \sum_{\vk, \vp}\sum_{s} |t_{\vk \vp}^{\rm n}|^2 \frac{\left|u^{\rm T}_{\vk s}v^{\rm B}_{\vp s} + v^{\rm T}_{\vk s} u^{\rm B}_{\vp s}\right|^2}{E^{\rm T}_{\vk s}+E^{\rm B}_{\vp s}}\\
    \Delta E_{\rm sf} &= - \sum_{\vk, \vp} \sum_{s}|t_{\vk \vp}^{\rm sf}|^2 \frac{\left|u^{\rm T}_{\vk s}v^{\rm B}_{\vp \bar{s}}  + v^{\rm T}_{\vk s} u^{\rm B}_{\vp \bar{s}}\right|^2}{E^{\rm T}_{\vk s}+E^{\rm B}_{\vp \bar{s}}}
\end{align}
with $E^l_{\vk s} = \sqrt{(\xi^l_{\vk s})^2 + |\Delta^l|^2}$ with the spin- and layer-dependent Bogoliubov transformation functions $u^{\rm {T,B}}_{\vk s}$ and $v^{\rm {T,B}}_{\vk s}$.
Using $(u_{\vk s}^l)^* v^{l}_{\vk s} = s|\Delta^l|\exp(i\phi_l)/E^l_{\vk s}$, we find for the phase-dependent contributions
\begin{align} \label{DeltaE}
\begin{split}
    E_{\rm J} = - \sum_{\vk, \vp} & \sum_{s}\Big[ |t_{\vk \vp}^{\rm n}|^2 \frac{|\Delta^{\rm T}\Delta^{\rm B}|}{E^{\rm T}_{\vk s}E^{\rm B}_{\vp s}(E^{\rm T}_{\vk s}+E^{\rm B}_{\vp s})}\\
    &- |t_{\vk \vp}^{\rm {sf}}|^2 \frac{|\Delta^{\rm T}\Delta^{\rm B}|}{E^{\rm T}_{\vk s}E^{\rm B}_{\vp \bar{s}}(E^{\rm T}_{\vk s}+E^{\rm B}_{\vp \bar{s}})}\Big] \cos(\phi_{\rm T} - \phi_{\rm B}).
    \end{split}
\end{align}
In the original discussion~\cite{kulik:1966}, momentum is not conserved in the tunneling process.  The sum is dominated by contributions close to the original Fermi surface, as there, $|u_{\vk s}^* v^{\phantom{*}}_{\vk s}| = |\Delta| / E_{\vk s} \propto 1$ and the energy denominator is the pairing gap. This gives rise to Eq.~\eqref{eq:Kulik}. 

We can now ask what happens if we consider the tunneling to be (almost) momentum conserving. In the usual case of tunneling through an oxide barrier, the common assumption is that momentum is not conserved. 
This is due to strong scattering at the interface and in the oxide barrier itself. 
In the case of stacked van der Waals materials, the situation is different. The interface between layers is smooth and if the intercalated molecules form an ordered array, there is little in-plane scattering. We note that momentum conservation applies even if hopping between molecular dimers is negligibly small.
For simplicity, we will proceed with the extreme case of perfect momentum conservation, considering the tunneling of a state with spin $s$ and momentum $\vk$ close to the original Fermi surface from the top layer to the bottom. Since the spin label in the dispersion is flipped between the layers, it is clear from Fig.~\ref{fig:bands}(a) that the final state with spin $s$ in the bottom layer is an excited state with energy given by the splitting between the red and blue bands, which is of order $\lambda$. 
This leads to a large energy denominator and  a small coherence factor in the first term in Eq.~\eqref{DeltaE}
resulting in an overall reduction by a factor $\Delta / \lambda \ll 1$.   
In contrast, the spin-flip contribution remains $O(1)$, as the final state can be near the Fermi surface.
To summarize, Eq.~\eqref{eq:Kulik} is replaced by
\begin{equation} \label{eq:Kulik2}
    J \propto (\alpha|t_{\rm n}|^2 - |t_{\rm sf}|^2),
\end{equation} 
where $\alpha \propto \Delta / \lambda \ll 1  $. This behavior is confirmed in Fig.~\ref{fig:bands}(b), where we used a momentum independent tunneling matrix element for both tunneling processes (For the quasi-momentum-conserving case, see App.~\ref{app:quasiconserved}). We now apply this equation to the 4Hb and the molecular-intercation cases.

\emph{Case 1: the 4Hb system---}%
The intermediate 1T layer started out as a  Mott insulator in a superlattice structure formed out of the  ``star-of-David'' charge density wave~\cite{wilson1975charge}.
The Mott insulator may be heavily depleted due to charge transfer. 
In the absence of disorder, momentum is conserved up to reciprocal superlattice vectors in tunneling. While van der Waals layers have negligible interface disorder, the star-of-David order gives a relatively small reciprocal lattice vector.  Importantly, it is known that there is a dilute distribution of local moments~\cite{shen2022coexistence,nayak2023}, which will give rise to  a finite  $t_{\rm sf}$. We assume the impurities to be sufficiently dilute and the superlattice scattering weak so that momentum is still conserved in the tunneling process. Furthermore, the pair-breaking effect of these impurities must not adversely affect the superconductivity in a significant way. It is therefore important to note that the electronic structure of the 1H layers is such that normal tunneling is strongly suppressed. A small amount of $t_{\rm sf}$ is sufficient for the second term in Eq.~\eqref{DeltaE} and ~\eqref{eq:Kulik2} to dominate, resulting in the negative Josephson coupling.

\emph{Case 2: Molecular intercalation---}%
 Eq.~\eqref{eq:Kulik2} applies equally well to the intercalation of chiral molecules. However, unlike the 4Hb case, we do not have sufficient understanding of the molecular system to provide an explicit mechanism for the requisite spin-flip scattering. Perhaps some kind of local moment is trapped near the contact point between the chiral molecule and the TaS$_2$ layer. We also note that our argument for the necessity of time-reversal breaking to form a $\pi$ junction (App.~\ref{app:TRS}) only refers to the formulation involving an effective tunneling Hamiltonian. In contrast, the mechanism described by Spivak and Kivelson \cite{spivak:1991} produces a $\pi$ junction without TRS breaking in a four-step process via a strongly-correlated intermediate state, which is outside of the effective tunneling picture. What we can state is that intercalation into 2H-TaS$_2$ has the special feature of strong suppression of spin-independent tunneling near the Fermi level. As a result, any other tunneling process involving either extrinsic defects or Coulomb correlation may dominate. While we do not have an explanation of why the control sample with achiral molecule does not show the effect, we note that only one achiral molecule, which is  quite different from the chiral molecule, was tested. It will be interesting to test other achiral molecules, especially ones with local moments, to see whether they show the $\pi$ phase shift.
Finally, we  point out that this system exhibits strong chiral-induced spin selectivity (CISS) which sets in at a finite temperature~\cite{qian2022chiral}. While the mechanism is not understood, some form of spin-dependent interaction in the barrier is generally taken as an essential starting point~\cite{evers2022}.

\emph{Discussion.}
In this paper, we have provided an alternative explanation for the observed $\pi$ phase shifts in the Little-Parks experiments in two systems, which does not postulate the appearance of a novel and exotic superconducting state. The common theme is that in both cases, the superconductivity resides in the 1H planes of the TMD superconductor. Furthermore, the breaking of inversion symmetry in the individual plane leads to spin-momentum locking of Ising type strongly suppressing non-spin-flip tunneling, leaving other spin-dependent processes that may change the sign of the inter-layer Josephson coupling.
 The sign change leads to the appearance of  $\pi$ phase shifts in the Little-Parks experiments about half the time on average. For the 4Hb case, we propose that the requisite spin-flip process can come from known paramagnetic impurities in the 1T layers.
 
\begin{figure}
    \centering
    \includegraphics{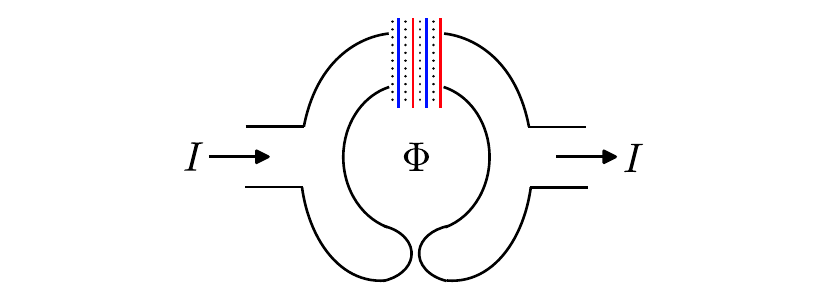}
    \caption{SQUID geometry to measure the negative interlayer Josephson coupling: For even number of layers, the SQUID enclosed a $\pi$ flux.}
    \label{fig:squid}
\end{figure}

While this paper offers an alternative explanation of the $\pi$ phase shift to that given in the original paper~\cite{almoalem:2022tmp}, we note that there have been several reports of intriguing findings in the 4Hb system~\cite{ribak:2020, nayak:2021, silber2022chiral, persky:2022}  that point to exotic pairing, particular of the time-reversal-breaking type. It is worth remarking that in our model, we can expect to find screw dislocations that either penetrate the sample or form dislocation loops in the bulk. 
These induce $\pi$-junctions and are unstable towards the formation of spontaneous current loops (clockwise or anti-clockwise), which may affect the  muon resonance relaxation rate~\cite{ribak:2020}. In order to settle the question definitively, it is a challenge for future experiments to directly detect the presence or absence of the $\pi$ phase in the Josephson coupling. In a vertical SQUID measurement, see Fig.~\ref{fig:squid}, the interference pattern will exhibit a $\pi$ shift in half of the samples if there is a negative interlayer Josephson coupling, while it is not expected in the case of a novel two-component order parameter. As a final note, we emphasize that our model has the general advantage of explaining why the superconductivity apparently survives in the dirty limit, as mentioned in the introduction. 

Finally, we put our finding in the context of other systems that have been proposed to show similar physics. In particular, it was argued that in almost decoupled layered materials, the even and odd stacking of $s$-wave order parameters is almost degenerate in energy, such that potentially even a weak perturbation, such as a magnetic field, could tip this balance. This has first been proposed for artificial heterostructures~\cite{yoshida:2012} and is believed to happen in the Ce-based heavy-fermion compound CeRh$_2$As$_2$~\cite{khim:2021}. Interestingly, the common scheme in these discussions is the local absence of inversion in the layers~\cite{fischer:2023}, with SOC driving the decoupling. In our scenario of spin-dependent tunneling between the layers, we argue that this tipping of the balance can happen without any external perturbation such as a magnetic field.

\emph{Acknowledgement.} We are grateful to Amit Kanigel, Avraham Klein, Rafael Fernandes, Igor Mazin, Gil Refael, and Manfred Sigrist for insightful discussions. Furthermore, we thank Erez Berg and Yuval Oreg for pointing out the importance of TRS breaking for the negative Josephson coupling. M.H.F. acknowledges financial support from the Swiss National Science Foundation (SNSF) through Division II (No. 207908). P.L. acknowledges the support by DOE office of Basic Sciences Grant No. DE-FG02-03ER46076. J.R. was supported by the Israeli Science
Foundation grant no. ISF-994/19.

\bibliography{ref}

\begin{thebibliography}{32}%
\makeatletter
\providecommand \@ifxundefined [1]{%
 \@ifx{#1\undefined}
}%
\providecommand \@ifnum [1]{%
 \ifnum #1\expandafter \@firstoftwo
 \else \expandafter \@secondoftwo
 \fi
}%
\providecommand \@ifx [1]{%
 \ifx #1\expandafter \@firstoftwo
 \else \expandafter \@secondoftwo
 \fi
}%
\providecommand \natexlab [1]{#1}%
\providecommand \enquote  [1]{``#1''}%
\providecommand \bibnamefont  [1]{#1}%
\providecommand \bibfnamefont [1]{#1}%
\providecommand \citenamefont [1]{#1}%
\providecommand \href@noop [0]{\@secondoftwo}%
\providecommand \href [0]{\begingroup \@sanitize@url \@href}%
\providecommand \@href[1]{\@@startlink{#1}\@@href}%
\providecommand \@@href[1]{\endgroup#1\@@endlink}%
\providecommand \@sanitize@url [0]{\catcode `\\12\catcode `\$12\catcode
  `\&12\catcode `\#12\catcode `\^12\catcode `\_12\catcode `\%12\relax}%
\providecommand \@@startlink[1]{}%
\providecommand \@@endlink[0]{}%
\providecommand \url  [0]{\begingroup\@sanitize@url \@url }%
\providecommand \@url [1]{\endgroup\@href {#1}{\urlprefix }}%
\providecommand \urlprefix  [0]{URL }%
\providecommand \Eprint [0]{\href }%
\providecommand \doibase [0]{https://doi.org/}%
\providecommand \selectlanguage [0]{\@gobble}%
\providecommand \bibinfo  [0]{\@secondoftwo}%
\providecommand \bibfield  [0]{\@secondoftwo}%
\providecommand \translation [1]{[#1]}%
\providecommand \BibitemOpen [0]{}%
\providecommand \bibitemStop [0]{}%
\providecommand \bibitemNoStop [0]{.\EOS\space}%
\providecommand \EOS [0]{\spacefactor3000\relax}%
\providecommand \BibitemShut  [1]{\csname bibitem#1\endcsname}%
\let\auto@bib@innerbib\@empty
%</preamble>
\bibitem [{\citenamefont {Geshkenbein}\ \emph {et~al.}(1987)\citenamefont
  {Geshkenbein}, \citenamefont {Larkin},\ and\ \citenamefont
  {Barone}}]{geshkenbein1987vortices}%
  \BibitemOpen
  \bibfield  {author} {\bibinfo {author} {\bibfnamefont {V.~B.}\ \bibnamefont
  {Geshkenbein}}, \bibinfo {author} {\bibfnamefont {A.~I.}\ \bibnamefont
  {Larkin}},\ and\ \bibinfo {author} {\bibfnamefont {A.}~\bibnamefont
  {Barone}},\ }\bibfield  {title} {\bibinfo {title} {Vortices with half
  magnetic flux quanta in ‘‘heavy-fermion’’superconductors},\
  }\href@noop {} {\bibfield  {journal} {\bibinfo  {journal} {Physical Review
  B}\ }\textbf {\bibinfo {volume} {36}},\ \bibinfo {pages} {235} (\bibinfo
  {year} {1987})}\BibitemShut {NoStop}%
\bibitem [{\citenamefont {Li}\ \emph {et~al.}(2019)\citenamefont {Li},
  \citenamefont {Xu}, \citenamefont {Lee}, \citenamefont {Chu},\ and\
  \citenamefont {Chien}}]{li:2019a}%
  \BibitemOpen
  \bibfield  {author} {\bibinfo {author} {\bibfnamefont {Y.}~\bibnamefont
  {Li}}, \bibinfo {author} {\bibfnamefont {X.}~\bibnamefont {Xu}}, \bibinfo
  {author} {\bibfnamefont {M.-H.}\ \bibnamefont {Lee}}, \bibinfo {author}
  {\bibfnamefont {M.-W.}\ \bibnamefont {Chu}},\ and\ \bibinfo {author}
  {\bibfnamefont {C.~L.}\ \bibnamefont {Chien}},\ }\bibfield  {title} {\bibinfo
  {title} {Observation of half-quantum flux in the unconventional
  superconductor bi$_2$ pd},\ }\href@noop {} {\bibfield  {journal} {\bibinfo
  {journal} {Science}\ }\textbf {\bibinfo {volume} {366}},\ \bibinfo {pages}
  {238} (\bibinfo {year} {2019})}\BibitemShut {NoStop}%
\bibitem [{\citenamefont {Xu}\ \emph {et~al.}(2020)\citenamefont {Xu},
  \citenamefont {Li},\ and\ \citenamefont {Chien}}]{xu:2020}%
  \BibitemOpen
  \bibfield  {author} {\bibinfo {author} {\bibfnamefont {X.}~\bibnamefont
  {Xu}}, \bibinfo {author} {\bibfnamefont {Y.}~\bibnamefont {Li}},\ and\
  \bibinfo {author} {\bibfnamefont {C.~L.}\ \bibnamefont {Chien}},\ }\bibfield
  {title} {\bibinfo {title} {Spin-triplet pairing state evidenced by
  half-quantum flux in a noncentrosymmetric superconductor},\ }\href
  {https://doi.org/10.1103/PhysRevLett.124.167001} {\bibfield  {journal}
  {\bibinfo  {journal} {Phys. Rev. Lett.}\ }\textbf {\bibinfo {volume} {124}},\
  \bibinfo {pages} {167001} (\bibinfo {year} {2020})}\BibitemShut {NoStop}%
\bibitem [{\citenamefont {Wan}\ \emph {et~al.}(2023)\citenamefont {Wan},
  \citenamefont {Qiu}, \citenamefont {Ren}, \citenamefont {Qian}, \citenamefont
  {Xu}, \citenamefont {Zhou}, \citenamefont {Zhou}, \citenamefont {Zhou},
  \citenamefont {Wang}, \citenamefont {Huang}, \citenamefont {Wang},\ and\
  \citenamefont {Duan}}]{wan:2023tmp}%
  \BibitemOpen
  \bibfield  {author} {\bibinfo {author} {\bibfnamefont {Z.}~\bibnamefont
  {Wan}}, \bibinfo {author} {\bibfnamefont {G.}~\bibnamefont {Qiu}}, \bibinfo
  {author} {\bibfnamefont {H.}~\bibnamefont {Ren}}, \bibinfo {author}
  {\bibfnamefont {Q.}~\bibnamefont {Qian}}, \bibinfo {author} {\bibfnamefont
  {D.}~\bibnamefont {Xu}}, \bibinfo {author} {\bibfnamefont {J.}~\bibnamefont
  {Zhou}}, \bibinfo {author} {\bibfnamefont {J.}~\bibnamefont {Zhou}}, \bibinfo
  {author} {\bibfnamefont {B.}~\bibnamefont {Zhou}}, \bibinfo {author}
  {\bibfnamefont {L.}~\bibnamefont {Wang}}, \bibinfo {author} {\bibfnamefont
  {Y.}~\bibnamefont {Huang}}, \bibinfo {author} {\bibfnamefont {K.~L.}\
  \bibnamefont {Wang}},\ and\ \bibinfo {author} {\bibfnamefont
  {X.}~\bibnamefont {Duan}},\ }\href@noop {} {\bibinfo {title} {Signatures of
  chiral superconductivity in chiral molecule intercalated tantalum disulfide}}
  (\bibinfo {year} {2023}),\ \Eprint {https://arxiv.org/abs/2302.05078}
  {arXiv:2302.05078 [cond-mat.supr-con]} \BibitemShut {NoStop}%
\bibitem [{\citenamefont {Almoalem}\ \emph {et~al.}(2022)\citenamefont
  {Almoalem}, \citenamefont {Feldman}, \citenamefont {Shlafman}, \citenamefont
  {Yaish}, \citenamefont {Fischer}, \citenamefont {Moshe}, \citenamefont
  {Ruhman},\ and\ \citenamefont {Kanigel}}]{almoalem:2022tmp}%
  \BibitemOpen
  \bibfield  {author} {\bibinfo {author} {\bibfnamefont {A.}~\bibnamefont
  {Almoalem}}, \bibinfo {author} {\bibfnamefont {I.}~\bibnamefont {Feldman}},
  \bibinfo {author} {\bibfnamefont {M.}~\bibnamefont {Shlafman}}, \bibinfo
  {author} {\bibfnamefont {Y.~E.}\ \bibnamefont {Yaish}}, \bibinfo {author}
  {\bibfnamefont {M.~H.}\ \bibnamefont {Fischer}}, \bibinfo {author}
  {\bibfnamefont {M.}~\bibnamefont {Moshe}}, \bibinfo {author} {\bibfnamefont
  {J.}~\bibnamefont {Ruhman}},\ and\ \bibinfo {author} {\bibfnamefont
  {A.}~\bibnamefont {Kanigel}},\ }\bibfield  {title} {\bibinfo {title}
  {Evidence of a two-component order parameter in 4hb-tas$_2$ in the
  little-parks effect},\ }\href@noop {} {\bibfield  {journal} {\bibinfo
  {journal} {arXiv preprint arXiv:2208.13798}\ } (\bibinfo {year}
  {2022})}\BibitemShut {NoStop}%
\bibitem [{\citenamefont {Nagata}\ \emph {et~al.}(1992)\citenamefont {Nagata},
  \citenamefont {Aochi}, \citenamefont {Abe}, \citenamefont {Ebisu},
  \citenamefont {Hagino}, \citenamefont {Seki},\ and\ \citenamefont
  {Tsutsumi}}]{nagata:1992}%
  \BibitemOpen
  \bibfield  {author} {\bibinfo {author} {\bibfnamefont {S.}~\bibnamefont
  {Nagata}}, \bibinfo {author} {\bibfnamefont {T.}~\bibnamefont {Aochi}},
  \bibinfo {author} {\bibfnamefont {T.}~\bibnamefont {Abe}}, \bibinfo {author}
  {\bibfnamefont {S.}~\bibnamefont {Ebisu}}, \bibinfo {author} {\bibfnamefont
  {T.}~\bibnamefont {Hagino}}, \bibinfo {author} {\bibfnamefont
  {Y.}~\bibnamefont {Seki}},\ and\ \bibinfo {author} {\bibfnamefont
  {K.}~\bibnamefont {Tsutsumi}},\ }\bibfield  {title} {\bibinfo {title}
  {Superconductivity in the layered compound 2h-tas$_2$},\ }\href@noop {}
  {\bibfield  {journal} {\bibinfo  {journal} {Journal of Physics and Chemistry
  of Solids}\ }\textbf {\bibinfo {volume} {53}},\ \bibinfo {pages} {1259}
  (\bibinfo {year} {1992})}\BibitemShut {NoStop}%
\bibitem [{\citenamefont {Navarro-Moratalla}\ \emph {et~al.}(2016)\citenamefont
  {Navarro-Moratalla}, \citenamefont {Island}, \citenamefont
  {Ma{\~n}as-Valero}, \citenamefont {Pinilla-Cienfuegos}, \citenamefont
  {Castellanos-Gomez}, \citenamefont {Quereda}, \citenamefont
  {Rubio-Bollinger}, \citenamefont {Chirolli}, \citenamefont
  {Silva-Guill{\'e}n}, \citenamefont {Agra{\"\i}t}, \citenamefont {Steele},
  \citenamefont {Guinea}, \citenamefont {van~der Zant},\ and\ \citenamefont
  {Coronado}}]{navarro-moratalla:2016}%
  \BibitemOpen
  \bibfield  {author} {\bibinfo {author} {\bibfnamefont {E.}~\bibnamefont
  {Navarro-Moratalla}}, \bibinfo {author} {\bibfnamefont {J.~O.}\ \bibnamefont
  {Island}}, \bibinfo {author} {\bibfnamefont {S.}~\bibnamefont
  {Ma{\~n}as-Valero}}, \bibinfo {author} {\bibfnamefont {E.}~\bibnamefont
  {Pinilla-Cienfuegos}}, \bibinfo {author} {\bibfnamefont {A.}~\bibnamefont
  {Castellanos-Gomez}}, \bibinfo {author} {\bibfnamefont {J.}~\bibnamefont
  {Quereda}}, \bibinfo {author} {\bibfnamefont {G.}~\bibnamefont
  {Rubio-Bollinger}}, \bibinfo {author} {\bibfnamefont {L.}~\bibnamefont
  {Chirolli}}, \bibinfo {author} {\bibfnamefont {J.~A.}\ \bibnamefont
  {Silva-Guill{\'e}n}}, \bibinfo {author} {\bibfnamefont {N.}~\bibnamefont
  {Agra{\"\i}t}}, \bibinfo {author} {\bibfnamefont {G.~A.}\ \bibnamefont
  {Steele}}, \bibinfo {author} {\bibfnamefont {F.}~\bibnamefont {Guinea}},
  \bibinfo {author} {\bibfnamefont {H.~S.~J.}\ \bibnamefont {van~der Zant}},\
  and\ \bibinfo {author} {\bibfnamefont {E.}~\bibnamefont {Coronado}},\
  }\bibfield  {title} {\bibinfo {title} {Enhanced superconductivity in
  atomically thin tas$_2$},\ }\href@noop {} {\bibfield  {journal} {\bibinfo
  {journal} {Nature Communications}\ }\textbf {\bibinfo {volume} {7}},\
  \bibinfo {pages} {11043} (\bibinfo {year} {2016})}\BibitemShut {NoStop}%
\bibitem [{\citenamefont {Bekaert}\ \emph {et~al.}(2020)\citenamefont
  {Bekaert}, \citenamefont {Khestanova}, \citenamefont {Hopkinson},
  \citenamefont {Birkbeck}, \citenamefont {Clark}, \citenamefont {Zhu},
  \citenamefont {Bandurin}, \citenamefont {Gorbachev}, \citenamefont
  {Fairclough}, \citenamefont {Zou}, \citenamefont {Hamer}, \citenamefont
  {Terry}, \citenamefont {Peters}, \citenamefont {Sanchez}, \citenamefont
  {Partoens}, \citenamefont {Haigh}, \citenamefont {Milo{\v s}evi{\'c}},\ and\
  \citenamefont {Grigorieva}}]{bekaert:2020}%
  \BibitemOpen
  \bibfield  {author} {\bibinfo {author} {\bibfnamefont {J.}~\bibnamefont
  {Bekaert}}, \bibinfo {author} {\bibfnamefont {E.}~\bibnamefont {Khestanova}},
  \bibinfo {author} {\bibfnamefont {D.~G.}\ \bibnamefont {Hopkinson}}, \bibinfo
  {author} {\bibfnamefont {J.}~\bibnamefont {Birkbeck}}, \bibinfo {author}
  {\bibfnamefont {N.}~\bibnamefont {Clark}}, \bibinfo {author} {\bibfnamefont
  {M.}~\bibnamefont {Zhu}}, \bibinfo {author} {\bibfnamefont {D.~A.}\
  \bibnamefont {Bandurin}}, \bibinfo {author} {\bibfnamefont {R.}~\bibnamefont
  {Gorbachev}}, \bibinfo {author} {\bibfnamefont {S.}~\bibnamefont
  {Fairclough}}, \bibinfo {author} {\bibfnamefont {Y.}~\bibnamefont {Zou}},
  \bibinfo {author} {\bibfnamefont {M.}~\bibnamefont {Hamer}}, \bibinfo
  {author} {\bibfnamefont {D.~J.}\ \bibnamefont {Terry}}, \bibinfo {author}
  {\bibfnamefont {J.~J.~P.}\ \bibnamefont {Peters}}, \bibinfo {author}
  {\bibfnamefont {A.~M.}\ \bibnamefont {Sanchez}}, \bibinfo {author}
  {\bibfnamefont {B.}~\bibnamefont {Partoens}}, \bibinfo {author}
  {\bibfnamefont {S.~J.}\ \bibnamefont {Haigh}}, \bibinfo {author}
  {\bibfnamefont {M.~V.}\ \bibnamefont {Milo{\v s}evi{\'c}}},\ and\ \bibinfo
  {author} {\bibfnamefont {I.~V.}\ \bibnamefont {Grigorieva}},\ }\bibfield
  {title} {\bibinfo {title} {Enhanced superconductivity in few-layer tas$_2$
  due to healing by oxygenation},\ }\href@noop {} {\bibfield  {journal}
  {\bibinfo  {journal} {Nano Letters}\ }\textbf {\bibinfo {volume} {20}},\
  \bibinfo {pages} {3808} (\bibinfo {year} {2020})}\BibitemShut {NoStop}%
\bibitem [{\citenamefont {Yang}\ \emph {et~al.}(2018)\citenamefont {Yang},
  \citenamefont {Fang}, \citenamefont {Fatemi}, \citenamefont {Ruhman},
  \citenamefont {Navarro-Moratalla}, \citenamefont {Watanabe}, \citenamefont
  {Taniguchi}, \citenamefont {Kaxiras},\ and\ \citenamefont
  {Jarillo-Herrero}}]{yang2018enhanced}%
  \BibitemOpen
  \bibfield  {author} {\bibinfo {author} {\bibfnamefont {Y.}~\bibnamefont
  {Yang}}, \bibinfo {author} {\bibfnamefont {S.}~\bibnamefont {Fang}}, \bibinfo
  {author} {\bibfnamefont {V.}~\bibnamefont {Fatemi}}, \bibinfo {author}
  {\bibfnamefont {J.}~\bibnamefont {Ruhman}}, \bibinfo {author} {\bibfnamefont
  {E.}~\bibnamefont {Navarro-Moratalla}}, \bibinfo {author} {\bibfnamefont
  {K.}~\bibnamefont {Watanabe}}, \bibinfo {author} {\bibfnamefont
  {T.}~\bibnamefont {Taniguchi}}, \bibinfo {author} {\bibfnamefont
  {E.}~\bibnamefont {Kaxiras}},\ and\ \bibinfo {author} {\bibfnamefont
  {P.}~\bibnamefont {Jarillo-Herrero}},\ }\bibfield  {title} {\bibinfo {title}
  {Enhanced superconductivity upon weakening of charge density wave transport
  in 2h-tas$_2$ in the two-dimensional limit},\ }\href@noop {} {\bibfield
  {journal} {\bibinfo  {journal} {Physical Review B}\ }\textbf {\bibinfo
  {volume} {98}},\ \bibinfo {pages} {035203} (\bibinfo {year}
  {2018})}\BibitemShut {NoStop}%
\bibitem [{Note1()}]{Note1}%
  \BibitemOpen
  \bibinfo {note} {The residual resistivity of 4Hb-TaS$_2$ is of the order of
  $\rho \approx 65 \mu \protect \rm {\Omega }\protect \,\protect \rm
  {cm}$~\cite {AmitPrivate}, which is 60 times greater than Sr$_2$RuO$_4$\cite
  {MackenzieExtremely}. Using Drude formula with a density $n=10^{19}$
  cm$^{-2}$ per layer and a rough estimate for the Fermi velocity $v_F =
  3\times 10^5$ m/s one reaches the conclusion that the mean-free path is of
  the order $\ell \approx 2$ nm. This is smaller than the coherence length
  which is of order $\xi \approx 20$ nm~\cite {ribak:2020}}\BibitemShut
  {NoStop}%
\bibitem [{Note2()}]{Note2}%
  \BibitemOpen
  \bibinfo {note} {Note that 0 and $\pi $ phase shifts are both allowed when
  time-reversal symmetry (TRS) is preserved, while broken TRS in principle
  allows for any phase.}\BibitemShut {Stop}%
\bibitem [{\citenamefont {Shiba}\ and\ \citenamefont
  {Soda}(1969)}]{shiba:1969}%
  \BibitemOpen
  \bibfield  {author} {\bibinfo {author} {\bibfnamefont {H.}~\bibnamefont
  {Shiba}}\ and\ \bibinfo {author} {\bibfnamefont {T.}~\bibnamefont {Soda}},\
  }\bibfield  {title} {\bibinfo {title} {{Superconducting Tunneling through the
  Barrier with Paramagnetic Impurities}},\ }\href
  {https://doi.org/10.1143/PTP.41.25} {\bibfield  {journal} {\bibinfo
  {journal} {Progress of Theoretical Physics}\ }\textbf {\bibinfo {volume}
  {41}},\ \bibinfo {pages} {25} (\bibinfo {year} {1969})}\BibitemShut {NoStop}%
\bibitem [{\citenamefont {Bulaevskii}\ \emph {et~al.}(1977)\citenamefont
  {Bulaevskii}, \citenamefont {Kuzii},\ and\ \citenamefont
  {Sobyanin}}]{bulaevskii:1977}%
  \BibitemOpen
  \bibfield  {author} {\bibinfo {author} {\bibfnamefont {L.~N.}\ \bibnamefont
  {Bulaevskii}}, \bibinfo {author} {\bibfnamefont {V.~V.}\ \bibnamefont
  {Kuzii}},\ and\ \bibinfo {author} {\bibfnamefont {A.~A.}\ \bibnamefont
  {Sobyanin}},\ }\bibfield  {title} {\bibinfo {title} {Superconducting system
  with weak coupling to the current in the ground state},\ }\bibfield
  {journal} {\bibinfo  {journal} {JETP Lett. (USSR) (Engl. Transl.); (United
  States)}\ }\textbf {\bibinfo {volume} {25}},\ \href
  {https://www.osti.gov/biblio/7316063} {} (\bibinfo {year} {1977})\BibitemShut
  {NoStop}%
\bibitem [{\citenamefont {Fischer}\ \emph {et~al.}(2023)\citenamefont
  {Fischer}, \citenamefont {Sigrist}, \citenamefont {Agterberg},\ and\
  \citenamefont {Yanase}}]{fischer:2023}%
  \BibitemOpen
  \bibfield  {author} {\bibinfo {author} {\bibfnamefont {M.~H.}\ \bibnamefont
  {Fischer}}, \bibinfo {author} {\bibfnamefont {M.}~\bibnamefont {Sigrist}},
  \bibinfo {author} {\bibfnamefont {D.~F.}\ \bibnamefont {Agterberg}},\ and\
  \bibinfo {author} {\bibfnamefont {Y.}~\bibnamefont {Yanase}},\ }\bibfield
  {title} {\bibinfo {title} {Superconductivity and local inversion-symmetry
  breaking},\ }\href@noop {} {\bibfield  {journal} {\bibinfo  {journal} {Annual
  Review of Condensed Matter Physics}\ }\textbf {\bibinfo {volume} {14}},\
  \bibinfo {pages} {153} (\bibinfo {year} {2023})}\BibitemShut {NoStop}%
\bibitem [{\citenamefont {Kulik}(1966)}]{kulik:1966}%
  \BibitemOpen
  \bibfield  {author} {\bibinfo {author} {\bibfnamefont {I.}~\bibnamefont
  {Kulik}},\ }\bibfield  {title} {\bibinfo {title} {Magnitude of the critical
  josephson tunnel current},\ }\href@noop {} {\bibfield  {journal} {\bibinfo
  {journal} {Soviet Journal of Experimental and Theoretical Physics}\ }\textbf
  {\bibinfo {volume} {22}},\ \bibinfo {pages} {841} (\bibinfo {year}
  {1966})}\BibitemShut {NoStop}%
\bibitem [{\citenamefont {Spivak}\ and\ \citenamefont
  {Kivelson}(1991)}]{spivak:1991}%
  \BibitemOpen
  \bibfield  {author} {\bibinfo {author} {\bibfnamefont {B.~I.}\ \bibnamefont
  {Spivak}}\ and\ \bibinfo {author} {\bibfnamefont {S.~A.}\ \bibnamefont
  {Kivelson}},\ }\bibfield  {title} {\bibinfo {title} {Negative local
  superfluid densities: The difference between dirty superconductors and dirty
  bose liquids},\ }\href@noop {} {\bibfield  {journal} {\bibinfo  {journal}
  {Phys. Rev. B}\ }\textbf {\bibinfo {volume} {43}},\ \bibinfo {pages} {3740}
  (\bibinfo {year} {1991})}\BibitemShut {NoStop}%
\bibitem [{\citenamefont {M\"ockli}\ and\ \citenamefont
  {Khodas}(2018)}]{moeckli:2018}%
  \BibitemOpen
  \bibfield  {author} {\bibinfo {author} {\bibfnamefont {D.}~\bibnamefont
  {M\"ockli}}\ and\ \bibinfo {author} {\bibfnamefont {M.}~\bibnamefont
  {Khodas}},\ }\bibfield  {title} {\bibinfo {title} {Robust parity-mixed
  superconductivity in disordered monolayer transition metal dichalcogenides},\
  }\href {https://doi.org/10.1103/PhysRevB.98.144518} {\bibfield  {journal}
  {\bibinfo  {journal} {Phys. Rev. B}\ }\textbf {\bibinfo {volume} {98}},\
  \bibinfo {pages} {144518} (\bibinfo {year} {2018})}\BibitemShut {NoStop}%
\bibitem [{\citenamefont {Liu}\ \emph {et~al.}(2013)\citenamefont {Liu},
  \citenamefont {Shan}, \citenamefont {Yao}, \citenamefont {Yao},\ and\
  \citenamefont {Xiao}}]{liu:2013}%
  \BibitemOpen
  \bibfield  {author} {\bibinfo {author} {\bibfnamefont {G.-B.}\ \bibnamefont
  {Liu}}, \bibinfo {author} {\bibfnamefont {W.-Y.}\ \bibnamefont {Shan}},
  \bibinfo {author} {\bibfnamefont {Y.}~\bibnamefont {Yao}}, \bibinfo {author}
  {\bibfnamefont {W.}~\bibnamefont {Yao}},\ and\ \bibinfo {author}
  {\bibfnamefont {D.}~\bibnamefont {Xiao}},\ }\bibfield  {title} {\bibinfo
  {title} {Three-band tight-binding model for monolayers of group-vib
  transition metal dichalcogenides},\ }\href
  {https://doi.org/10.1103/PhysRevB.88.085433} {\bibfield  {journal} {\bibinfo
  {journal} {Phys. Rev. B}\ }\textbf {\bibinfo {volume} {88}},\ \bibinfo
  {pages} {085433} (\bibinfo {year} {2013})}\BibitemShut {NoStop}%
\bibitem [{\citenamefont {de~la Barrera}\ \emph {et~al.}(2018)\citenamefont
  {de~la Barrera}, \citenamefont {Sinko}, \citenamefont {Gopalan},
  \citenamefont {Sivadas}, \citenamefont {Seyler}, \citenamefont {Watanabe},
  \citenamefont {Taniguchi}, \citenamefont {Tsen}, \citenamefont {Xu},
  \citenamefont {Xiao},\ and\ \citenamefont {Hunt}}]{barrera:2018}%
  \BibitemOpen
  \bibfield  {author} {\bibinfo {author} {\bibfnamefont {S.~C.}\ \bibnamefont
  {de~la Barrera}}, \bibinfo {author} {\bibfnamefont {M.~R.}\ \bibnamefont
  {Sinko}}, \bibinfo {author} {\bibfnamefont {D.~P.}\ \bibnamefont {Gopalan}},
  \bibinfo {author} {\bibfnamefont {N.}~\bibnamefont {Sivadas}}, \bibinfo
  {author} {\bibfnamefont {K.~L.}\ \bibnamefont {Seyler}}, \bibinfo {author}
  {\bibfnamefont {K.}~\bibnamefont {Watanabe}}, \bibinfo {author}
  {\bibfnamefont {T.}~\bibnamefont {Taniguchi}}, \bibinfo {author}
  {\bibfnamefont {A.~W.}\ \bibnamefont {Tsen}}, \bibinfo {author}
  {\bibfnamefont {X.}~\bibnamefont {Xu}}, \bibinfo {author} {\bibfnamefont
  {D.}~\bibnamefont {Xiao}},\ and\ \bibinfo {author} {\bibfnamefont {B.~M.}\
  \bibnamefont {Hunt}},\ }\bibfield  {title} {\bibinfo {title} {Tuning ising
  superconductivity with layer and spin--orbit coupling in two-dimensional
  transition-metal dichalcogenides},\ }\href@noop {} {\bibfield  {journal}
  {\bibinfo  {journal} {Nature Communications}\ }\textbf {\bibinfo {volume}
  {9}},\ \bibinfo {pages} {1427} (\bibinfo {year} {2018})}\BibitemShut
  {NoStop}%
\bibitem [{\citenamefont {Wilson}\ \emph {et~al.}(1975)\citenamefont {Wilson},
  \citenamefont {Di~Salvo},\ and\ \citenamefont {Mahajan}}]{wilson1975charge}%
  \BibitemOpen
  \bibfield  {author} {\bibinfo {author} {\bibfnamefont {J.~A.}\ \bibnamefont
  {Wilson}}, \bibinfo {author} {\bibfnamefont {F.}~\bibnamefont {Di~Salvo}},\
  and\ \bibinfo {author} {\bibfnamefont {S.}~\bibnamefont {Mahajan}},\
  }\bibfield  {title} {\bibinfo {title} {Charge-density waves and superlattices
  in the metallic layered transition metal dichalcogenides},\ }\href@noop {}
  {\bibfield  {journal} {\bibinfo  {journal} {Advances in Physics}\ }\textbf
  {\bibinfo {volume} {24}},\ \bibinfo {pages} {117} (\bibinfo {year}
  {1975})}\BibitemShut {NoStop}%
\bibitem [{\citenamefont {Shen}\ \emph {et~al.}(2022)\citenamefont {Shen},
  \citenamefont {Qin}, \citenamefont {Gao}, \citenamefont {Wen}, \citenamefont
  {Wang}, \citenamefont {Wang}, \citenamefont {Li}, \citenamefont {Luo},
  \citenamefont {Lu}, \citenamefont {Sun} \emph
  {et~al.}}]{shen2022coexistence}%
  \BibitemOpen
  \bibfield  {author} {\bibinfo {author} {\bibfnamefont {S.}~\bibnamefont
  {Shen}}, \bibinfo {author} {\bibfnamefont {T.}~\bibnamefont {Qin}}, \bibinfo
  {author} {\bibfnamefont {J.}~\bibnamefont {Gao}}, \bibinfo {author}
  {\bibfnamefont {C.}~\bibnamefont {Wen}}, \bibinfo {author} {\bibfnamefont
  {J.}~\bibnamefont {Wang}}, \bibinfo {author} {\bibfnamefont {W.}~\bibnamefont
  {Wang}}, \bibinfo {author} {\bibfnamefont {J.}~\bibnamefont {Li}}, \bibinfo
  {author} {\bibfnamefont {X.}~\bibnamefont {Luo}}, \bibinfo {author}
  {\bibfnamefont {W.}~\bibnamefont {Lu}}, \bibinfo {author} {\bibfnamefont
  {Y.}~\bibnamefont {Sun}}, \emph {et~al.},\ }\bibfield  {title} {\bibinfo
  {title} {Coexistence of quasi-two-dimensional superconductivity and tunable
  kondo lattice in a van der waals superconductor},\ }\href@noop {} {\bibfield
  {journal} {\bibinfo  {journal} {Chinese Physics Letters}\ }\textbf {\bibinfo
  {volume} {39}},\ \bibinfo {pages} {077401} (\bibinfo {year}
  {2022})}\BibitemShut {NoStop}%
\bibitem [{\citenamefont {Nayak}\ \emph {et~al.}(2023)\citenamefont {Nayak},
  \citenamefont {Steinbok}, \citenamefont {Roet}, \citenamefont {Koo},
  \citenamefont {Feldman}, \citenamefont {Almoalem}, \citenamefont {Kanigel},
  \citenamefont {Yan}, \citenamefont {Rosch}, \citenamefont {Avraham} \emph
  {et~al.}}]{nayak2023}%
  \BibitemOpen
  \bibfield  {author} {\bibinfo {author} {\bibfnamefont {A.~K.}\ \bibnamefont
  {Nayak}}, \bibinfo {author} {\bibfnamefont {A.}~\bibnamefont {Steinbok}},
  \bibinfo {author} {\bibfnamefont {Y.}~\bibnamefont {Roet}}, \bibinfo {author}
  {\bibfnamefont {J.}~\bibnamefont {Koo}}, \bibinfo {author} {\bibfnamefont
  {I.}~\bibnamefont {Feldman}}, \bibinfo {author} {\bibfnamefont
  {A.}~\bibnamefont {Almoalem}}, \bibinfo {author} {\bibfnamefont
  {A.}~\bibnamefont {Kanigel}}, \bibinfo {author} {\bibfnamefont
  {B.}~\bibnamefont {Yan}}, \bibinfo {author} {\bibfnamefont {A.}~\bibnamefont
  {Rosch}}, \bibinfo {author} {\bibfnamefont {N.}~\bibnamefont {Avraham}},
  \emph {et~al.},\ }\bibfield  {title} {\bibinfo {title} {First order quantum
  phase transition in the hybrid metal-mott insulator transition metal
  dichalcogenide 4hb-tas$_2$},\ }\href@noop {} {\bibfield  {journal} {\bibinfo
  {journal} {arXiv preprint arXiv:2303.01447}\ } (\bibinfo {year}
  {2023})}\BibitemShut {NoStop}%
\bibitem [{\citenamefont {Qian}\ \emph {et~al.}(2022)\citenamefont {Qian},
  \citenamefont {Ren}, \citenamefont {Zhou}, \citenamefont {Wan}, \citenamefont
  {Zhou}, \citenamefont {Yan}, \citenamefont {Cai}, \citenamefont {Wang},
  \citenamefont {Li}, \citenamefont {Sofer} \emph {et~al.}}]{qian2022chiral}%
  \BibitemOpen
  \bibfield  {author} {\bibinfo {author} {\bibfnamefont {Q.}~\bibnamefont
  {Qian}}, \bibinfo {author} {\bibfnamefont {H.}~\bibnamefont {Ren}}, \bibinfo
  {author} {\bibfnamefont {J.}~\bibnamefont {Zhou}}, \bibinfo {author}
  {\bibfnamefont {Z.}~\bibnamefont {Wan}}, \bibinfo {author} {\bibfnamefont
  {J.}~\bibnamefont {Zhou}}, \bibinfo {author} {\bibfnamefont {X.}~\bibnamefont
  {Yan}}, \bibinfo {author} {\bibfnamefont {J.}~\bibnamefont {Cai}}, \bibinfo
  {author} {\bibfnamefont {P.}~\bibnamefont {Wang}}, \bibinfo {author}
  {\bibfnamefont {B.}~\bibnamefont {Li}}, \bibinfo {author} {\bibfnamefont
  {Z.}~\bibnamefont {Sofer}}, \emph {et~al.},\ }\bibfield  {title} {\bibinfo
  {title} {Chiral molecular intercalation superlattices},\ }\href@noop {}
  {\bibfield  {journal} {\bibinfo  {journal} {Nature}\ }\textbf {\bibinfo
  {volume} {606}},\ \bibinfo {pages} {902} (\bibinfo {year}
  {2022})}\BibitemShut {NoStop}%
\bibitem [{\citenamefont {Evers}\ \emph {et~al.}(2022)\citenamefont {Evers},
  \citenamefont {Aharony}, \citenamefont {Bar-Gill}, \citenamefont
  {Entin-Wohlman}, \citenamefont {Hedeg{\aa}rd}, \citenamefont {Hod},
  \citenamefont {Jelinek}, \citenamefont {Kamieniarz}, \citenamefont
  {Lemeshko}, \citenamefont {Michaeli} \emph {et~al.}}]{evers2022}%
  \BibitemOpen
  \bibfield  {author} {\bibinfo {author} {\bibfnamefont {F.}~\bibnamefont
  {Evers}}, \bibinfo {author} {\bibfnamefont {A.}~\bibnamefont {Aharony}},
  \bibinfo {author} {\bibfnamefont {N.}~\bibnamefont {Bar-Gill}}, \bibinfo
  {author} {\bibfnamefont {O.}~\bibnamefont {Entin-Wohlman}}, \bibinfo {author}
  {\bibfnamefont {P.}~\bibnamefont {Hedeg{\aa}rd}}, \bibinfo {author}
  {\bibfnamefont {O.}~\bibnamefont {Hod}}, \bibinfo {author} {\bibfnamefont
  {P.}~\bibnamefont {Jelinek}}, \bibinfo {author} {\bibfnamefont
  {G.}~\bibnamefont {Kamieniarz}}, \bibinfo {author} {\bibfnamefont
  {M.}~\bibnamefont {Lemeshko}}, \bibinfo {author} {\bibfnamefont
  {K.}~\bibnamefont {Michaeli}}, \emph {et~al.},\ }\bibfield  {title} {\bibinfo
  {title} {Theory of chirality induced spin selectivity: Progress and
  challenges},\ }\href@noop {} {\bibfield  {journal} {\bibinfo  {journal}
  {Advanced Materials}\ }\textbf {\bibinfo {volume} {34}},\ \bibinfo {pages}
  {2106629} (\bibinfo {year} {2022})}\BibitemShut {NoStop}%
\bibitem [{\citenamefont {Ribak}\ \emph {et~al.}(2020)\citenamefont {Ribak},
  \citenamefont {Skiff}, \citenamefont {Mograbi}, \citenamefont {Rout},
  \citenamefont {Fischer}, \citenamefont {Ruhman}, \citenamefont {Chashka},
  \citenamefont {Dagan},\ and\ \citenamefont {Kanigel}}]{ribak:2020}%
  \BibitemOpen
  \bibfield  {author} {\bibinfo {author} {\bibfnamefont {A.}~\bibnamefont
  {Ribak}}, \bibinfo {author} {\bibfnamefont {R.~M.}\ \bibnamefont {Skiff}},
  \bibinfo {author} {\bibfnamefont {M.}~\bibnamefont {Mograbi}}, \bibinfo
  {author} {\bibfnamefont {P.~K.}\ \bibnamefont {Rout}}, \bibinfo {author}
  {\bibfnamefont {M.~H.}\ \bibnamefont {Fischer}}, \bibinfo {author}
  {\bibfnamefont {J.}~\bibnamefont {Ruhman}}, \bibinfo {author} {\bibfnamefont
  {K.}~\bibnamefont {Chashka}}, \bibinfo {author} {\bibfnamefont
  {Y.}~\bibnamefont {Dagan}},\ and\ \bibinfo {author} {\bibfnamefont
  {A.}~\bibnamefont {Kanigel}},\ }\bibfield  {title} {\bibinfo {title} {Chiral
  superconductivity in the alternate stacking compound 4hb-tas$_2$},\ }\href
  {https://doi.org/10.1126/sciadv.aax9480} {\bibfield  {journal} {\bibinfo
  {journal} {Science Advances}\ }\textbf {\bibinfo {volume} {6}},\ \bibinfo
  {pages} {aax9480} (\bibinfo {year} {2020})}\BibitemShut {NoStop}%
\bibitem [{\citenamefont {Nayak}\ \emph {et~al.}(2021)\citenamefont {Nayak},
  \citenamefont {Steinbok}, \citenamefont {Roet}, \citenamefont {Koo},
  \citenamefont {Margalit}, \citenamefont {Feldman}, \citenamefont {Almoalem},
  \citenamefont {Kanigel}, \citenamefont {Fiete}, \citenamefont {Yan},
  \citenamefont {Oreg}, \citenamefont {Avraham},\ and\ \citenamefont
  {Beidenkopf}}]{nayak:2021}%
  \BibitemOpen
  \bibfield  {author} {\bibinfo {author} {\bibfnamefont {A.~K.}\ \bibnamefont
  {Nayak}}, \bibinfo {author} {\bibfnamefont {A.}~\bibnamefont {Steinbok}},
  \bibinfo {author} {\bibfnamefont {Y.}~\bibnamefont {Roet}}, \bibinfo {author}
  {\bibfnamefont {J.}~\bibnamefont {Koo}}, \bibinfo {author} {\bibfnamefont
  {G.}~\bibnamefont {Margalit}}, \bibinfo {author} {\bibfnamefont
  {I.}~\bibnamefont {Feldman}}, \bibinfo {author} {\bibfnamefont
  {A.}~\bibnamefont {Almoalem}}, \bibinfo {author} {\bibfnamefont
  {A.}~\bibnamefont {Kanigel}}, \bibinfo {author} {\bibfnamefont {G.~A.}\
  \bibnamefont {Fiete}}, \bibinfo {author} {\bibfnamefont {B.}~\bibnamefont
  {Yan}}, \bibinfo {author} {\bibfnamefont {Y.}~\bibnamefont {Oreg}}, \bibinfo
  {author} {\bibfnamefont {N.}~\bibnamefont {Avraham}},\ and\ \bibinfo {author}
  {\bibfnamefont {H.}~\bibnamefont {Beidenkopf}},\ }\bibfield  {title}
  {\bibinfo {title} {Evidence of topological boundary modes with topological
  nodal-point superconductivity},\ }\href@noop {} {\bibfield  {journal}
  {\bibinfo  {journal} {Nature Physics}\ }\textbf {\bibinfo {volume} {17}},\
  \bibinfo {pages} {1413} (\bibinfo {year} {2021})}\BibitemShut {NoStop}%
\bibitem [{\citenamefont {Silber}\ \emph {et~al.}(2022)\citenamefont {Silber},
  \citenamefont {Mathimalar}, \citenamefont {Mangel}, \citenamefont {Green},
  \citenamefont {Avraham}, \citenamefont {Beidenkopf}, \citenamefont {Feldman},
  \citenamefont {Kanigel}, \citenamefont {Klein}, \citenamefont {Goldstein},
  \citenamefont {Banerjee}, \citenamefont {Sela},\ and\ \citenamefont
  {Dagan}}]{silber2022chiral}%
  \BibitemOpen
  \bibfield  {author} {\bibinfo {author} {\bibfnamefont {I.}~\bibnamefont
  {Silber}}, \bibinfo {author} {\bibfnamefont {S.}~\bibnamefont {Mathimalar}},
  \bibinfo {author} {\bibfnamefont {I.}~\bibnamefont {Mangel}}, \bibinfo
  {author} {\bibfnamefont {O.}~\bibnamefont {Green}}, \bibinfo {author}
  {\bibfnamefont {N.}~\bibnamefont {Avraham}}, \bibinfo {author} {\bibfnamefont
  {H.}~\bibnamefont {Beidenkopf}}, \bibinfo {author} {\bibfnamefont
  {I.}~\bibnamefont {Feldman}}, \bibinfo {author} {\bibfnamefont
  {A.}~\bibnamefont {Kanigel}}, \bibinfo {author} {\bibfnamefont
  {A.}~\bibnamefont {Klein}}, \bibinfo {author} {\bibfnamefont
  {M.}~\bibnamefont {Goldstein}}, \bibinfo {author} {\bibfnamefont
  {A.}~\bibnamefont {Banerjee}}, \bibinfo {author} {\bibfnamefont
  {E.}~\bibnamefont {Sela}},\ and\ \bibinfo {author} {\bibfnamefont
  {Y.}~\bibnamefont {Dagan}},\ }\href@noop {} {\bibinfo {title} {Chiral to
  nematic crossover in the superconducting state of 4hb-tas$_2$}} (\bibinfo
  {year} {2022}),\ \Eprint {https://arxiv.org/abs/2208.14442} {arXiv:2208.14442
  [cond-mat.supr-con]} \BibitemShut {NoStop}%
\bibitem [{\citenamefont {Persky}\ \emph {et~al.}(2022)\citenamefont {Persky},
  \citenamefont {Bj{\o}rlig}, \citenamefont {Feldman}, \citenamefont
  {Almoalem}, \citenamefont {Altman}, \citenamefont {Berg}, \citenamefont
  {Kimchi}, \citenamefont {Ruhman}, \citenamefont {Kanigel},\ and\
  \citenamefont {Kalisky}}]{persky:2022}%
  \BibitemOpen
  \bibfield  {author} {\bibinfo {author} {\bibfnamefont {E.}~\bibnamefont
  {Persky}}, \bibinfo {author} {\bibfnamefont {A.~V.}\ \bibnamefont
  {Bj{\o}rlig}}, \bibinfo {author} {\bibfnamefont {I.}~\bibnamefont {Feldman}},
  \bibinfo {author} {\bibfnamefont {A.}~\bibnamefont {Almoalem}}, \bibinfo
  {author} {\bibfnamefont {E.}~\bibnamefont {Altman}}, \bibinfo {author}
  {\bibfnamefont {E.}~\bibnamefont {Berg}}, \bibinfo {author} {\bibfnamefont
  {I.}~\bibnamefont {Kimchi}}, \bibinfo {author} {\bibfnamefont
  {J.}~\bibnamefont {Ruhman}}, \bibinfo {author} {\bibfnamefont
  {A.}~\bibnamefont {Kanigel}},\ and\ \bibinfo {author} {\bibfnamefont
  {B.}~\bibnamefont {Kalisky}},\ }\bibfield  {title} {\bibinfo {title}
  {Magnetic memory and spontaneous vortices in a van der waals
  superconductor},\ }\href@noop {} {\bibfield  {journal} {\bibinfo  {journal}
  {Nature}\ }\textbf {\bibinfo {volume} {607}},\ \bibinfo {pages} {692}
  (\bibinfo {year} {2022})}\BibitemShut {NoStop}%
\bibitem [{\citenamefont {Yoshida}\ \emph {et~al.}(2012)\citenamefont
  {Yoshida}, \citenamefont {Sigrist},\ and\ \citenamefont
  {Yanase}}]{yoshida:2012}%
  \BibitemOpen
  \bibfield  {author} {\bibinfo {author} {\bibfnamefont {T.}~\bibnamefont
  {Yoshida}}, \bibinfo {author} {\bibfnamefont {M.}~\bibnamefont {Sigrist}},\
  and\ \bibinfo {author} {\bibfnamefont {Y.}~\bibnamefont {Yanase}},\
  }\bibfield  {title} {\bibinfo {title} {Pair-density wave states through
  spin-orbit coupling in multilayer superconductors},\ }\href@noop {}
  {\bibfield  {journal} {\bibinfo  {journal} {Phys. Rev. B}\ }\textbf {\bibinfo
  {volume} {86}},\ \bibinfo {pages} {134514} (\bibinfo {year}
  {2012})}\BibitemShut {NoStop}%
\bibitem [{\citenamefont {Khim}\ \emph {et~al.}(2021)\citenamefont {Khim},
  \citenamefont {Landaeta}, \citenamefont {Banda}, \citenamefont {Bannor},
  \citenamefont {Brando}, \citenamefont {Brydon}, \citenamefont {Hafner},
  \citenamefont {K{\"u}chler}, \citenamefont {Cardoso-Gil}, \citenamefont
  {Stockert}, \citenamefont {Mackenzie}, \citenamefont {Agterberg},
  \citenamefont {Geibel},\ and\ \citenamefont {Hassinger}}]{khim:2021}%
  \BibitemOpen
  \bibfield  {author} {\bibinfo {author} {\bibfnamefont {S.}~\bibnamefont
  {Khim}}, \bibinfo {author} {\bibfnamefont {J.~F.}\ \bibnamefont {Landaeta}},
  \bibinfo {author} {\bibfnamefont {J.}~\bibnamefont {Banda}}, \bibinfo
  {author} {\bibfnamefont {N.}~\bibnamefont {Bannor}}, \bibinfo {author}
  {\bibfnamefont {M.}~\bibnamefont {Brando}}, \bibinfo {author} {\bibfnamefont
  {P.~M.~R.}\ \bibnamefont {Brydon}}, \bibinfo {author} {\bibfnamefont
  {D.}~\bibnamefont {Hafner}}, \bibinfo {author} {\bibfnamefont
  {R.}~\bibnamefont {K{\"u}chler}}, \bibinfo {author} {\bibfnamefont
  {R.}~\bibnamefont {Cardoso-Gil}}, \bibinfo {author} {\bibfnamefont
  {U.}~\bibnamefont {Stockert}}, \bibinfo {author} {\bibfnamefont {A.~P.}\
  \bibnamefont {Mackenzie}}, \bibinfo {author} {\bibfnamefont {D.~F.}\
  \bibnamefont {Agterberg}}, \bibinfo {author} {\bibfnamefont {C.}~\bibnamefont
  {Geibel}},\ and\ \bibinfo {author} {\bibfnamefont {E.}~\bibnamefont
  {Hassinger}},\ }\bibfield  {title} {\bibinfo {title} {Field-induced
  transition within the superconducting state of cerh$_2$as$_2$},\ }\href@noop
  {} {\bibfield  {journal} {\bibinfo  {journal} {Science}\ }\textbf {\bibinfo
  {volume} {373}},\ \bibinfo {pages} {1012} (\bibinfo {year}
  {2021})}\BibitemShut {NoStop}%
\bibitem [{\citenamefont {Kanigel}()}]{AmitPrivate}%
  \BibitemOpen
  \bibfield  {author} {\bibinfo {author} {\bibfnamefont {A.}~\bibnamefont
  {Kanigel}},\ }\href@noop {} {}\bibinfo {howpublished} {private
  communication}\BibitemShut {NoStop}%
\bibitem [{\citenamefont {Mackenzie}\ \emph {et~al.}(1998)\citenamefont
  {Mackenzie}, \citenamefont {Haselwimmer}, \citenamefont {Tyler},
  \citenamefont {Lonzarich}, \citenamefont {Mori}, \citenamefont {Nishizaki},\
  and\ \citenamefont {Maeno}}]{MackenzieExtremely}%
  \BibitemOpen
  \bibfield  {author} {\bibinfo {author} {\bibfnamefont {A.~P.}\ \bibnamefont
  {Mackenzie}}, \bibinfo {author} {\bibfnamefont {R.~K.~W.}\ \bibnamefont
  {Haselwimmer}}, \bibinfo {author} {\bibfnamefont {A.~W.}\ \bibnamefont
  {Tyler}}, \bibinfo {author} {\bibfnamefont {G.~G.}\ \bibnamefont
  {Lonzarich}}, \bibinfo {author} {\bibfnamefont {Y.}~\bibnamefont {Mori}},
  \bibinfo {author} {\bibfnamefont {S.}~\bibnamefont {Nishizaki}},\ and\
  \bibinfo {author} {\bibfnamefont {Y.}~\bibnamefont {Maeno}},\ }\bibfield
  {title} {\bibinfo {title} {Extremely strong dependence of superconductivity
  on disorder in ${\mathrm{sr}}_{2}{\mathrm{ruo}}_{4}$},\ }\href
  {https://doi.org/10.1103/PhysRevLett.80.161} {\bibfield  {journal} {\bibinfo
  {journal} {Phys. Rev. Lett.}\ }\textbf {\bibinfo {volume} {80}},\ \bibinfo
  {pages} {161} (\bibinfo {year} {1998})}\BibitemShut {NoStop}%
\end{thebibliography}%

\newpage

\appendix

\begin{widetext}

\section{Ginzburg-Landau}\label{app:GL}
In this section, we introduce a Ginzburg-Landau formulation of the free energy for the two-layer structure and discuss the Little-Parks effect in this setup. Importantly, we use cyllindrical coordinates and treat the system as discrete in the $z$ direction, using a layer index $l$, and continuous in the plane, with radial coordinate $r$ and angle $\varphi$. In each layer $l$, the free energy density can be written as
\begin{equation}
    f_l[\eta] = \alpha(T-T_c^0) |\eta_l(\vrr)|^2 + \frac{\beta}{2} |\eta_l(\vrr)|^4 + \gamma |D_{\parallel} \eta_l(\vrr)|^2 + \frac{J}2 |\eta_l(\vrr) - \eta_{l+1}(\vrr)|^2,
\end{equation}
where $\eta_l(\vrr)$ is the order parameter in layer $l$ at (2D) position $\vrr$; $D_{\parallel} = (-i \hbar \vec{\nabla} + 2e \vec{A})_{\parallel}$ is the in-plane component of the covariant derivative; $\alpha$, $\beta$, $\gamma$ are phenomenological paramters; $T_{\rm c}^0$ is the `bare' critical temperature; and $J$ is the Josephson coupling between the layers. Note that this coupling yields the interlayer coupling of the main text with $J|\eta|^2 \rightarrow J$ up to a constant factor in the free energy. In the following, we compare the two cases of constant sign between the layers, $\eta^{0}_{l}(\vrr) = \eta(\vrr)$ and alternating sign, in other words $\eta^{\pi}_l(\vrr) = (-1)^l \eta(\vrr)$. Furthermore, we consider an annular sample consisting of $L$ layers and inner and outer radii $R_1$ and $R_2$, respectively.

Without a screw dislocation and no magnetic field, the interlayer term does not enter the constant-sign solution, such that $T_{\rm c}$ is unchanged, $T_{\rm c} = T_{\rm c}^0$. For the alternating sign, the critical temperature is shifted by $2J/\alpha$, $T^\pi_{\rm c} = T_{\rm c}^0 - 2J/\alpha$. For a negative coupling $J<0$, the solution $\eta^\pi_l(\vrr)$ thus has a higher critical temperature and is favored. 

If we add a screw dislocation, layer $l$ becomes layer $l+1$ after one rotation (we only consider this case here, as all other cases follow from the two cases of no and single screw dislocation). Put differently, a rotation of the angle $\varphi$ by $2\pi$, $\varphi \mapsto \varphi + 2\pi$ is equivalent to a translation in the $z$ direction by one layer, $l \mapsto l+1$. The ansatz for the solution stays the same for the constant-sign solution, but changes for the solution $\eta^\pi_l(\vrr)$. First, we consider a given point in the annulus, around which the phase in-plane is approximately constant, while the inter-layer coupling enforces a sign-changing structure in the out-of-plane direction. In other words, we assume a solution of the form
\begin{equation}
    \tilde{\eta}_l^\pi (r, \varphi=0) = |\eta| e^{i\pi l}.
\end{equation}
Note that the origin of this angle is arbitrary and does not influence the discussion. In order to smoothly connect to this solution when going around the annulus, in other words when $\varphi\mapsto\varphi + 2\pi$, the phase must wind, such that we make the ansatz
\begin{equation}
    \tilde{\eta}_l^\pi (r, \varphi) = |\eta| e^{i(\varphi /2 + \pi l)}.
\end{equation}
For such a solution, the rotation and translation by one layer are indeed equivalent. Note that the phase winding required to smoothly connect to the next layer after one rotation for this solution costs energy. In particular, the gradient term now reads
\begin{equation}
    \gamma |-i\vec{\nabla} \tilde{\eta}_l(r, \varphi)|^2 = \frac{\gamma}{4r^2} |\eta|^2.
\end{equation}
After integration over the sample, the free energy to second order yields
\begin{equation}
    F[\eta] = \sum_l \int r dr d\varphi f_l[\eta] = \alpha \pi (R_2^2 - R_1^2) L \Big[T-T_c^0 + \frac{2J}{\alpha} + \frac{\gamma}{2 \alpha (R_2^2 - R_1^2)L} \log(\frac{R_2}{R_1})\Big]   |\eta|^2.
\end{equation}
The critical temperature now reads
\begin{equation}
    \tilde{T}^\pi_{\rm c} = T_{\rm c}^0 - \frac{2J}{\alpha} + \frac{\gamma}{2 \alpha (R_2^2 - R_1^2) L } \log(\frac{R_1}{R_2})
\end{equation}
and is reduced by the cost of a half-quantum vortex. For a thin annulus, $R_1 = R_2(1-\delta)$ with $\delta\ll1$, we find for the critical temperature
\begin{equation}
    \tilde{T}^\pi_{\rm c} \approx T_{\rm c}^0 - \frac{2J}{\alpha} - \frac{\gamma}{4 \alpha R_2^2},
\end{equation}
such that for large enough samples and $J<0$, the solution with alternating sign has still the highest critical temperature.

Two comments are in order at this point: (1) Unlike the argument of a single-valued wave function in a doubly-connected geometry used for fluxoid quantization, our argument is an argument about energetics. As such, we compare the energy of two different ansatzes, namely a sign-preserving and a sign-changing one. As noted, we find
the sign-changing solution to be energetically favored, causing the ring to  realize a $\pi$ phase shift. (2) The two solutions we consider, which realize either a $0$ or a $\pi$ ring, are the only two solutions that are eigenfunctions of the translation operator and conserve time-reversal symmetry. While one could consider a different phase structure, in other words quasimomentum $k_z$ and total phase $\varphi k_z / (2\pi) + l k_z$, these order-parameter ansatzes are never energetically favored at zero field and the resulting trapped flux would not be $0$ or $\pi$ anymore, a situation that breaks time-reversal symmetry without an applied field.

Finally, we add a magnetic field out of plane in the gauge with $\vec{A} = (H r/2)  \hat{e}_\varphi$. Using the ansatz
\begin{equation}
    \tilde{\eta}^\pi_l (r, \varphi) = |\eta| e^{i[(2n + 1) \varphi /2 + \pi l]}
\end{equation}
yields the (covariant) gradient term
\begin{equation}
    \gamma |D_{\parallel} \eta^\pi_l(r, \varphi)|^2 = \gamma\Big(\frac{2n+1}{2r} - e H r \Big)^2 |\eta|^2,
\end{equation}
such that the critical temperature is found by choosing $n$ in order to maximize 
\begin{equation}
    T_{\rm c} = T_{\rm c}^0 - \frac{2J}{\alpha} - \frac{\gamma}{2 \alpha L (R_2^2 - R_1^2)}\int_{R_1}^{R_2} \frac{dr}{r} \Big(\frac{2n+1}{2} - \frac{\Phi}{\Phi_0}\frac{r^2}{R_1^2}\Big)^2.
\end{equation}
In this last equation, we have introduced the flux through the annular disk, $\Phi = \pi R_1^2 H$ and $\Phi_0=2e/h$ is the flux quantum.
Indeed, we find the expected $T_{\rm c}$ oscillations with a maximum at zero field for the standard $s$-wave case, but the $T_{\rm c}$ is minimum for the alternating-sign solution, as a field corresponding to half a flux quantum exactly cancels the cost for the phase winding.

\section{Absence of $\pi$ junction in the single particle formulation of tunneling with time-reversal symmetry.}\label{app:TRS}
\begin{figure}[tt]
\centering
\includegraphics{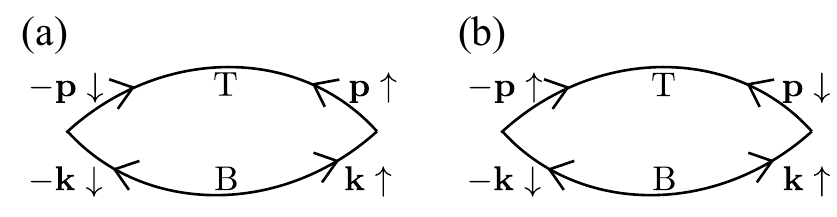}
\caption{Diagrams that contribute to the Josephson energy from (a) non-spin-flip tunneling and (b) spin-flip tunneling.}
\label{fig:diagram}
\end{figure}
In this section, we discuss the condition for obtaining a negative Josephson energy using a single-particle formulation. We assume a general spin-flip tunneling Hamiltonian
\begin{equation} \label{eq:app_hopping}
    \HH_{\rm tun} = \sum_{\vk,\vp}\ \{[T_{\vk \vp}^{\downarrow\uparrow} \; c^\dag_{\vk,{\rm T},\downarrow } c^{\pdag}_{\vp,{\rm B}, \uparrow} + T_{\vk \vp}^{\uparrow \downarrow} \; c^\dag_{\vk,{\rm T},\uparrow} c^{\pdag}_{\vp,{\rm B}, \downarrow}]+{\rm h.c.}\}.
\end{equation}
We calculate the Josephson energy using a diagrammatic technique using the anomalous Green function $F$ in Matsubara space for the superconducting state. With non-spin-flip hopping with the matrix element  $T_{\vk \vp}^n$, we find the energy from the diagram in Fig.~\ref{fig:diagram}(a)
\begin{equation} \label{normal}
E^n \propto  -\sum_{\omega_m} \int d \vp d\vk \ [T_{-\vk -\vp}^n T_{\vk \vp}^n F_{\rm T,\uparrow \downarrow}(\vk,\omega_m) F^*_{\rm B,\downarrow \uparrow}(\vp,\omega_m) +{\rm h.c.}].
\end{equation}
On the other hand, with spin-flip scattering we find from the diagram in Fig.~\ref{fig:diagram}(b)
\begin{equation} \label{spinflip}
E^s \propto  -\sum_{\omega_m} \int  d \vp d\vk \ [T_{-\vk -\vp}^{\uparrow \downarrow} T_{\vk \vp}^{\downarrow \uparrow} F_{\rm T,\downarrow \uparrow}(\vk,\omega_m) F^*_{\rm B,\downarrow \uparrow}(\vp,\omega_m) +{\rm h.c.}].
\end{equation}
The Josephson energy $E_J=-(E^n + E^s)$.
Since $F_{\rm T,\uparrow \downarrow}(\vk,\omega_m)=-F_{\rm T,\downarrow \uparrow}(\vk,\omega_m)=-\Delta_{\rm T}/(\omega_m^2+|\Delta_{\rm T}|^2+\xi_k ^2)$, this gives a potential 
sign change. However, in the presence of time reversal symmetry, we have 
\begin{equation} \label{time}
     T_{\vk \vp}^{\downarrow \uparrow} = -T_{-\vk -\vp}^{\uparrow \downarrow *}.
\end{equation}
As a result, Eq.~\eqref{spinflip} becomes
\begin{equation*} \label{spinflip3}
    E^s \propto  -\sum_{\omega_m} \int d \vp d\vk \ [ | T_{\vk \vp}^{\downarrow \uparrow}|^2 F_{\rm T,\uparrow \downarrow}(\vk,\omega_m) F^*_{\rm B,\downarrow \uparrow}(\vp,\omega_m) +{\rm h.c.}],
\end{equation*}
which has the same sign as for non-spin-flip scattering given by Eq.~\eqref{normal}. The two sign changes have cancelled each other. This is a general proof that in the presence of time-reversal-symmetry breaking, the Josephson energy is always positive in the single-particle treatment of tunneling. This result disagrees with a statement of Kulik \cite{kulik:1966}, who claimed that the spin-flip scattering he considered could be due to SOC, which does not break time reversal. We were not able to follow his definition of the spin flip term. In order to get a sign reversal, we need either tunneling terms which break time reversal, such as scattering from a set of random moments \cite{shiba:1969,bulaevskii:1977}, or a many-body treatment of correlation effect as done by Spivak and Kivelson. \cite{spivak:1991}

\section{Effekt of (quasi-) momentum-conserving tunneling}\label{app:quasiconserved}
In order to study the effect of (quasi) momentum conservation in the tunneling matrix element in the main text, we use a Lorentzian envelope to model the matrix elements
\begin{equation}
    t_{\vk\vp}^{\nu} = t_{\nu} \frac{1}{\pi} \frac{\Gamma/2}{|\vk - \vp|^2 + (\Gamma/2)^2}
\end{equation}
with $\nu = \textrm{n}, \textrm{sf}$ for the spin-conserving and spin-flip tunneling, respectively. With this simplified form, we can evaluate $\alpha$ in Eq.~\eqref{eq:Kulik2} directly on a lattice for a typical band structure for 1H layers combining to a 2H structure, as shown in the main text~\cite{moeckli:2018, liu:2013}. Figure~\ref{fig:alpha} shows the result for different gap magnitudes as a function of $\xi = \sqrt{3}/(2\pi \Gamma)$, the characteristic (real-space) length scale connected to momentum conservation.

\begin{figure}
    \centering
    \includegraphics{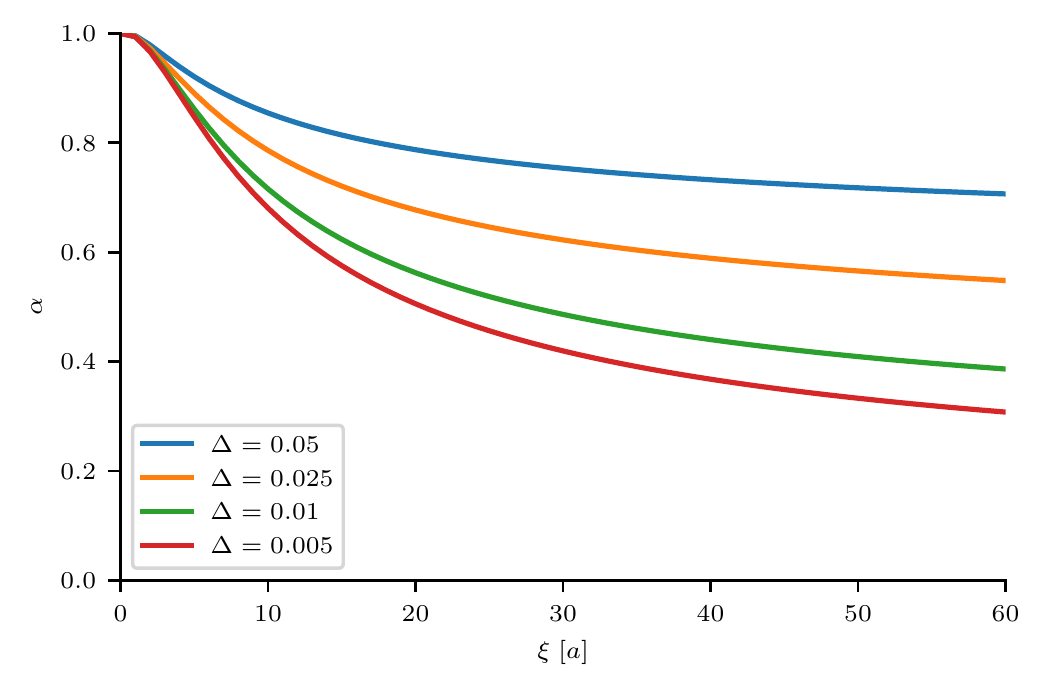}
    \caption{Weight of the spin-conserving tunneling compared to the spin-flip tunneling for a Lorentzian envelope for the tunneling matrix element. For the calculations, we used the parameterization of Refs~\cite{moeckli:2018, liu:2013} with $\lambda \approx 0.07$eV.}
    \label{fig:alpha}
\end{figure}
\end{widetext}

\end{document}